\documentclass[final,1p,times]{elsarticle} % do not change
\usepackage{graphicx} % do not change
\usepackage{amssymb} % do not change
\usepackage{amsthm} % do not change
\usepackage{lineno} % do not change

\journal{Nuclear Physics A} % do not change
\begin{document} % do not change

\begin{frontmatter} % do not change

%% QM09Author: please enter your  
%% Title, author and address info here; please do not use footnotes

% Your Title - please modify
\title{Heavy Ion Physics with CMS}

% Principle author, and co-authors - please modify
\author{Olga Kodolova$^{a}$ and Michael Murray$^{b}$ for the CMS collaboration}

% Address - please modify
% note that if you have authors from several institutions, we recommend
% labelling these [a], [b], [c] etc.
\address[a]{Moscow State University, % label [b]
119991,Leninskie Gory, 1/2, Moscow, Russia, kodolova@mail.cern.ch}
\address[b]{University of Kansas, % label [a]
Lawrence KS 66045-7582, USA,  mjmurray@ku.edu} 

\begin{abstract} % do not change
We present the capabilities of the CMS experiment to explore the heavy-ion physics program offered by the CERN Large Hadron Collider (LHC). 
The prime goal of this research is to test the fundamental theory of the strong interaction (QCD) in extreme conditions of temperature, density and parton momentum fraction by colliding nuclei at energies of $\sqrt{s_{_{NN}}}$ = 5.5 TeV. 
This presentation will give the overview of the potential of the CMS to carry out a full set of representative Pb-Pb measurements both in ''soft'' and ''hard'' regimes.
Measurements include ``bulk'' observables -- charged hadron multiplicity, low $p_{\rm T}$ inclusive hadron identified spectra and elliptic flow -- which provide information on the collective properties of the system; as well as perturbative processes -- such as quarkonia, heavy-quarks, jets, $\gamma$-jet, and high $p_{\rm T}$ hadrons --- which yield ``tomographic'' information of the hottest and densest phases of the reaction.
\end{abstract} % do not change

\end{frontmatter} % do not change

%% QM09: we keep linenumbers at least for initial version
%%\linenumbers % do not change

%% start of main text - please modify. Below is a sub-set (single section) 
%% of an earlier proceedings that show how one can handle references 
%% and figures etc.
%%\section{}\label{}

\section{Understanding the strong force under extreme conditions}

The study of the strong interaction (QCD) in extreme conditions of temperature and
density has been the driving force for experiments from the Bevalac to the the Large Hadron Collider.  In the last decade the four RHIC experiments have produced beautiful evidence that in the energy range $\sqrt{s_{NN}}=63-200$~GeV  a strongly interacting quark gluon liquid is produced \cite{Whitepapers}.  The scaling of the elliptic flow with quark number, the suppression of fast quarks in the medium  are clears signals of this but a great wealth of other evidence is shown in the proceedings of this conference.   At both SPS and RHIC energies 
the suppression of the  J/$\psi$ resonance suggests that we have created a very high temperature system ~\cite{NA38,jpsi_rhic_1,satz}. 
In addition there is evidence, from work at forward rapidities that at small parton momentum fraction (low-x) the  initial state of the nuclei may be a sheet of gluons, the color glass condensate \cite{Arsene:2004ux,Kharzeev:2002pc}. 
    The LHC plans to collide Pb nuclei at $\sqrt{s_{NN}}=5.5$~TeV which is 28 times higher than 
the highest energy available at RHIC. 
We expect the initial state to be 
dominated by  saturated parton distribution with relevant range of parton momentum fraction {\it x} as low as 
$10^{-5}$ and  a characteristic saturation momentum, $Q_{s}^{2}\simeq 5-10$ GeV$^{2}$ \cite{colorglas1}. The collisions should produce 
copious  hard probes such as jets, high-$p_T$ hadrons, heavy-quarks, quarkonia and  large yields 
of the weakly interacting perturbative probes (direct photons, dileptons, Z$^0$ and 
W$^{\pm}$ bosons)~\cite{hiaddendum}. This paper will concentrate on our best guess of the physics at the LHC with an emphasis on early measurements.  However the great strength of CMS is that it is a generic detector for heavy ions well suited to discovering the completely unexpected.

\section{The CMS detector}
\begin{figure}[ht]
\centering
\includegraphics[scale=0.35]{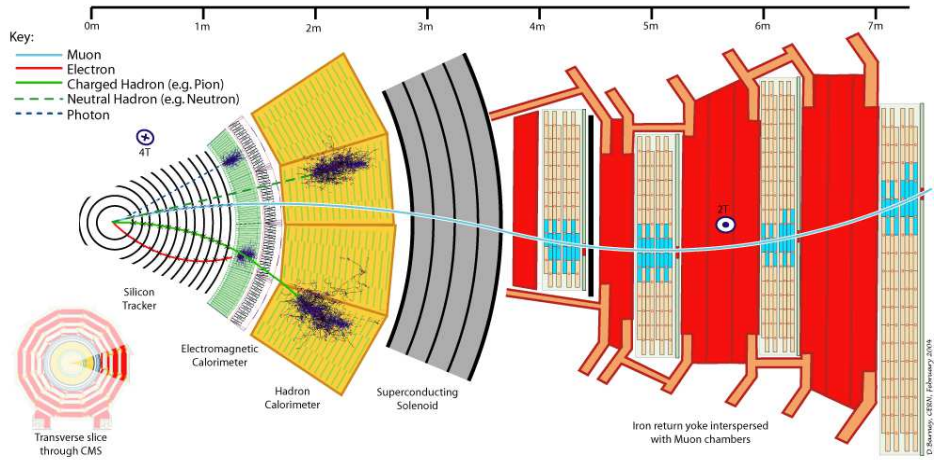}
\caption[]{(color online) CMS detector (slice view).\label{CMSdet}}
\end{figure}

%A detailed description of the Compact Muon Solenoid (CMS) experiment can be found elsewhere~\cite{JINST}. 
The central feature of the CMS apparatus is a  3.8T superconducting solenoid, of 6~m internal diameter.  Within 
the field volume there are the silicon pixel and strip tracker, the crystal electromagnetic calorimeter 
(ECAL) and the brass-scintillator hadronic calorimeter (HCAL).  Besides the barrel and endcap calorimeters ($|\eta|<3$), CMS has extensive forward calorimetry, HF ($3<|\eta|<5.2$), CASTOR ($5.3<|\eta|<6.6$) and Zero Degree ($|\eta|>8.3$) calorimeters. Muons are measured in gaseous chambers embedded in the iron return yoke.
A full description  of the experiment can be found elsewhere~\cite{JINST}. A slice through CMS is shown in Fig.~\ref{CMSdet}.

Near mid-rapidity ($|\eta|<2.5$) charged particles are tracked by three layers of silicon pixel detectors, made of 66~million $100\times150$~$\mu$m$^2$ pixels, followed by ten microstrip layers, with strips of pitch between 80 and 180~$\mu$m.
The silicon tracker provides the vertex position with $\sim$\,15~$\mu$m accuracy.
The ECAL has an energy resolution of better than 0.5\,\% 
above 100~GeV.  The HCAL, when combined with the ECAL, measures jets with a
resolution $\Delta E/E \approx 100\,\% / \sqrt{E} \oplus 5\,\%$.  The
calorimeter cells are grouped in projective towers, of granularity $\Delta \eta \times \Delta \phi = 0.087\times0.087$ at central rapidities and $0.175\times0.175$ at forward rapidities. Muons are measured in the pseudorapidity window $|\eta|< 2.4$, with detection planes made of three technologies: Drift Tubes, Cathode Strip Chambers, and Resistive Plate Chambers.  Matching the muons to
the tracks measured in the silicon tracker results in a transverse momentum resolution between 1 and 5\,\%, for $p_{\rm T}$ values up to 1~TeV/$c$.
The good momentum resolution of the  tracker allows us to clearly resolve the $\Upsilon$--family.

The first level (L1) of the CMS trigger system,  uses information from the calorimeters and muon detectors to select  the most interesting events (only one bunch crossing in 1000 in pp collisions). 
It is composed of custom hardware processors and takes less than 1~$\mu$s to reach a decision. 
 The High Level Trigger (HLT) processor farm further decreases the event rate from 100~kHz to
100~Hz, before data storage.  For Pb-Pb runs the low collision rate (8 kHz)  together with the fast L1
trigger will allow us to send almost all the events triggered by the L1 MinBias  trigger to HLT-farm and provide full reconstruction of events in real time.

\section{Bulk ("hydro") measurements in A-A collisions}
\begin{figure}[ht]
\centering
\includegraphics[scale=0.5,angle=-90]{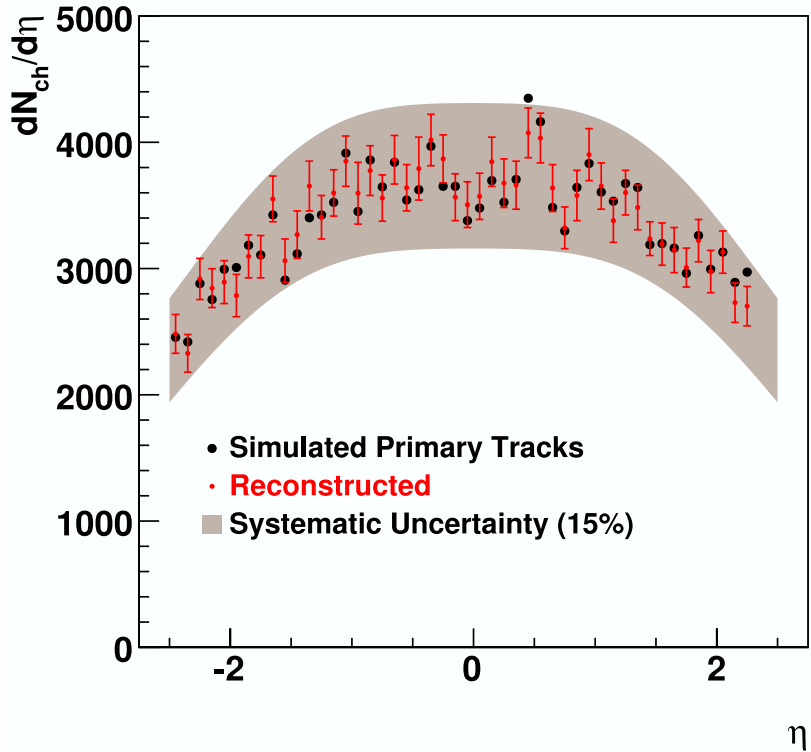}
\caption[]{(color online) Comparison of the original distribution of primary simulated
tracks (black) to the estimate obtained
from  reconstructed hits in layer 1 of the pixels 
(smaller)  for one central Pb-Pb event ~\cite{hiaddendum}. The grey band
indicates a somewhat conservative systematic uncertainty. 
\label{hit_cms}}
\end{figure}

%\subsection{Charged particle multiplicity}
The charged particle multiplicity per unit of rapidity at mid-rapidity is related to the  
entropy density in the collisions and fixes the global properties of the produced medium.  The unexpectedly low multiplicities seen at RHIC have lent support to the color glass picture and it will be interesting to see if this model works at LHC energies.  
%>>>Seems old: MJM 
%Extrapolations from SPS to RHIC energies gave essential overestimation relative to the measured multiplicity 
%at RHIC. This evidence gives rise to  the Color Glass Condensate (CGC) approaches which effectively take 
%into account a reduced initial number of scattering centers in PDFs and reproduces results from RHIC. 
%The expected hadron multiplicities at midrapidity in the frame of CGC model is much lower ($dN/d\eta|_{\eta=0}=$2000) 
%than the $dN/d\eta|_{\eta=0}=$8000 predictions before RHIC results.
CMS is planning to make a first day measurement of the charged particle multiplicities by two methods:
1) hit counting in the pixels using  a  {\it dE/dx} cut and 2) tracklets with a vertex constraint. Figure~\ref{hit_cms} shows that for one event we can accurately reconstruct $dN/d\eta$ using the  hit counting technique. 

%\subsection{Low-$p_T$ hadron spectra}

\begin{figure}[ht]
\centering
    \includegraphics[scale=0.3]{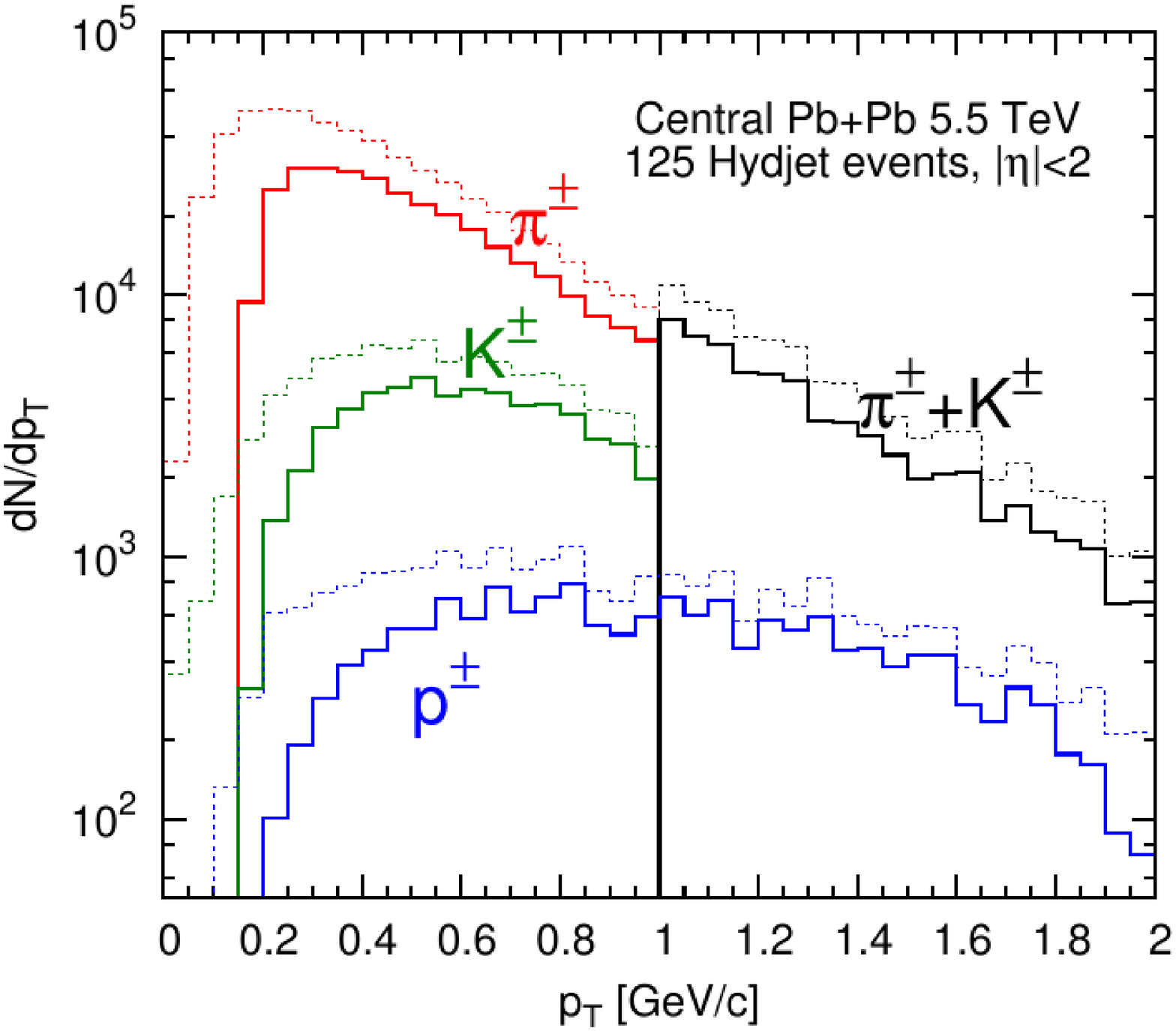}
    \caption{(color online) Low-$p_T$ spectra of generated hadrons (dotted line) and the reconstructed
    hadrons (solid lines). 
%    The simulation was done for central Pb-Pb events at $\sqrt{S_{NN}}=$5.5 TeV.
    \label{cms_lowpt}}
\end{figure}

Measurements of hadron momentum spectra and ratios at low $p_T$ are an important tool to determine the amount of
collective radial flow and the thermal and chemical conditions of the system at  freeze-out.
CMS has developed a special low $p_T$ tracking algorithm based only on the pixels. This allows us to
identify particles by comparing the  energy loss, {\it dE/dx}, and the momentum of the track. Inclusive hadron spectra can be measured from $p\simeq$400~MeV/c up to $p\simeq$1 GeV/c for pions and kaons and up to $p\simeq$2 GeV/c for protons (Fig.~\ref{cms_lowpt}).  

\begin{figure}[hbtp]
\centering
    \resizebox{0.4\textwidth}{!}{\includegraphics{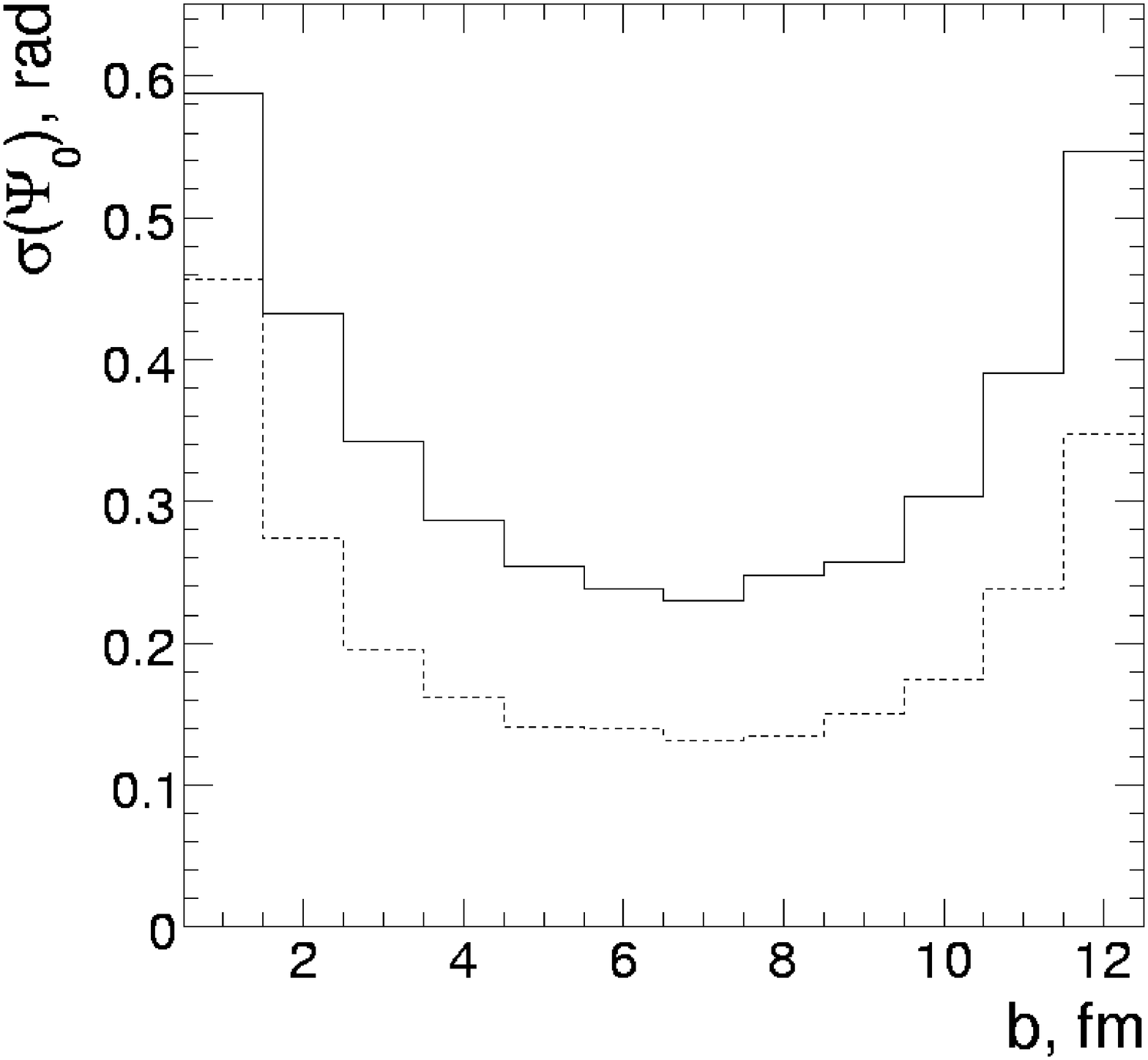}}
    \hfill
    \resizebox{0.49\textwidth}{!}{\includegraphics{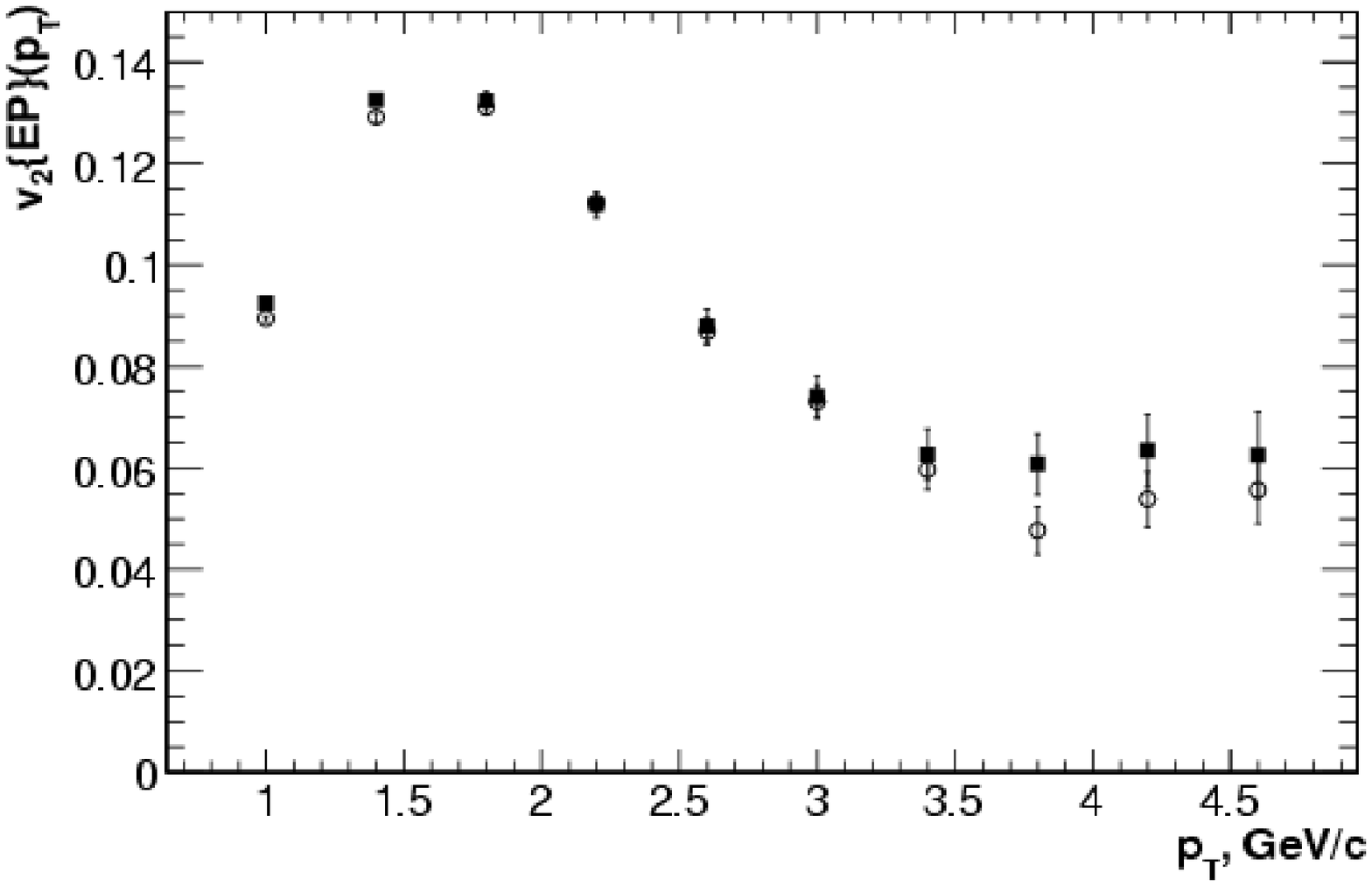}}
    \parbox[t]{0.47\textwidth}{\caption{Event plane resolution, $\sigma (\Psi _0)$,  as a function
    of impact parameter in Pb+Pb collisions with $N_0$ (b = 0 fm) = 58000 (solid
    histogram) and 84000 (dashed histogram) total particle multiplicities.}\label{evplane}}
    \hfill
    \parbox[t]{0.47\textwidth}{\caption{Reconstructed $v_2$ parameter on the $p_T$ 
                                    of track for CMS. Elliptic flow was included in HYDJET.~\cite{hydjet}
        }
    \label{cms_v2}}
\end{figure}

%\subsection{Elliptic flow}
Unless the two lead nuclei collide head on the overlap region will have an  elliptical shape. 
For a liquid, this initial space anisotropy is  translated into
a final elliptical asymmetry in momentum space. However for a gas any anisotropy should be much weaker. 
The elliptic flow parameter, $v_2$
is the  strength of the second harmonic of the the azimuthal
distribution of hadrons with respect to the reaction plane. Comparing the experimental  $v_2$
with hydrodynamical calculations will show us how close the matter is  fully thermalized perfect fluid 
close ~\cite{hiaddendum}.
   CMS will measure $v_2$ both by reconstructing the 
   reaction plane 
  and by using multiparticle correlators or cummulants. The 
resolution of the event plane determination for CMS is shown in Fig.~\ref{evplane} while  Fig.~\ref{cms_v2} 
shows our ability to measure $v_2$ as a function of $p_T$. 

\begin{figure}[hbtp]
\centering
    \resizebox{0.49\textwidth}{!}{\includegraphics{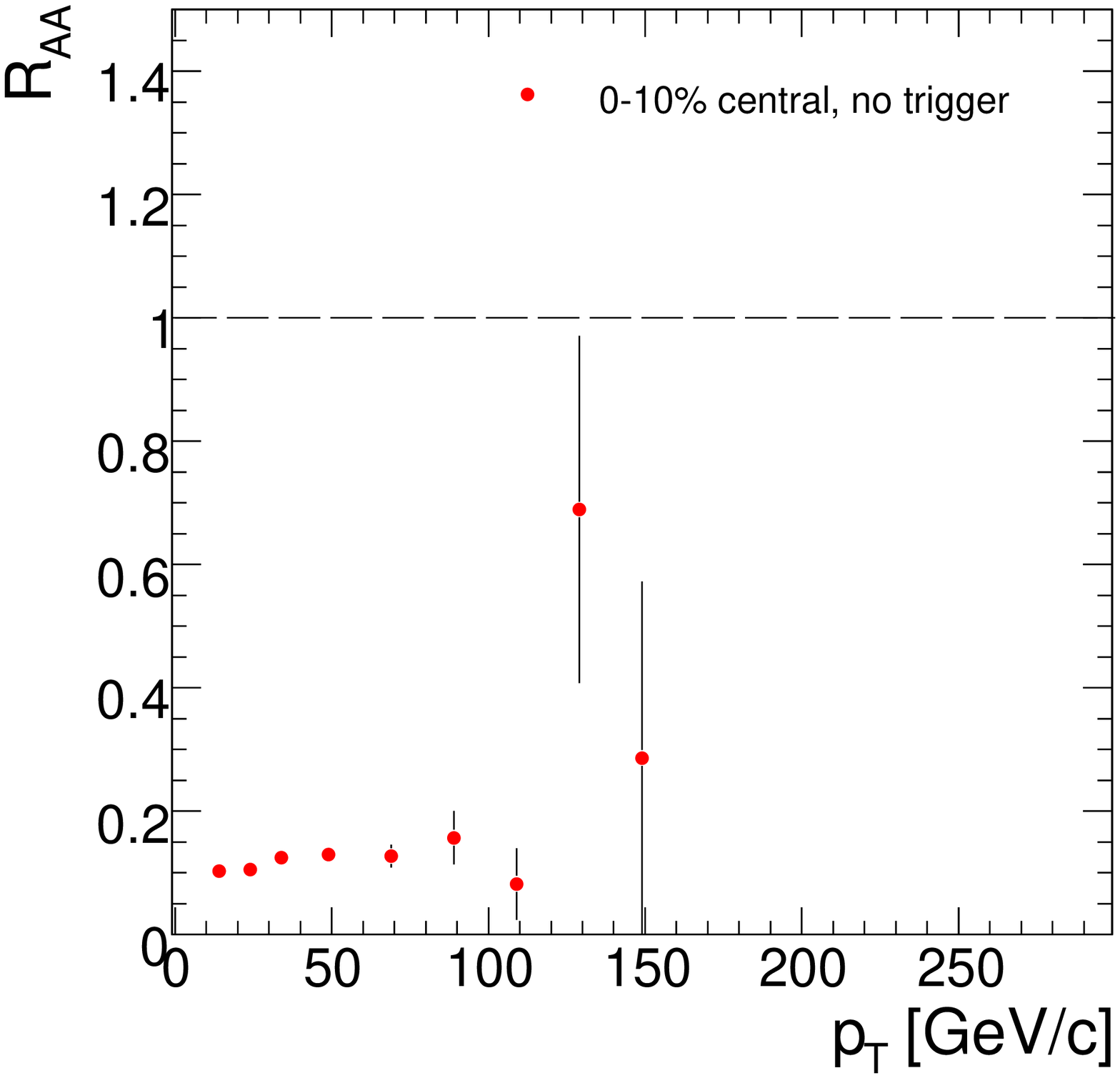}}
    \hfill
    \resizebox{0.49\textwidth}{!}{\includegraphics{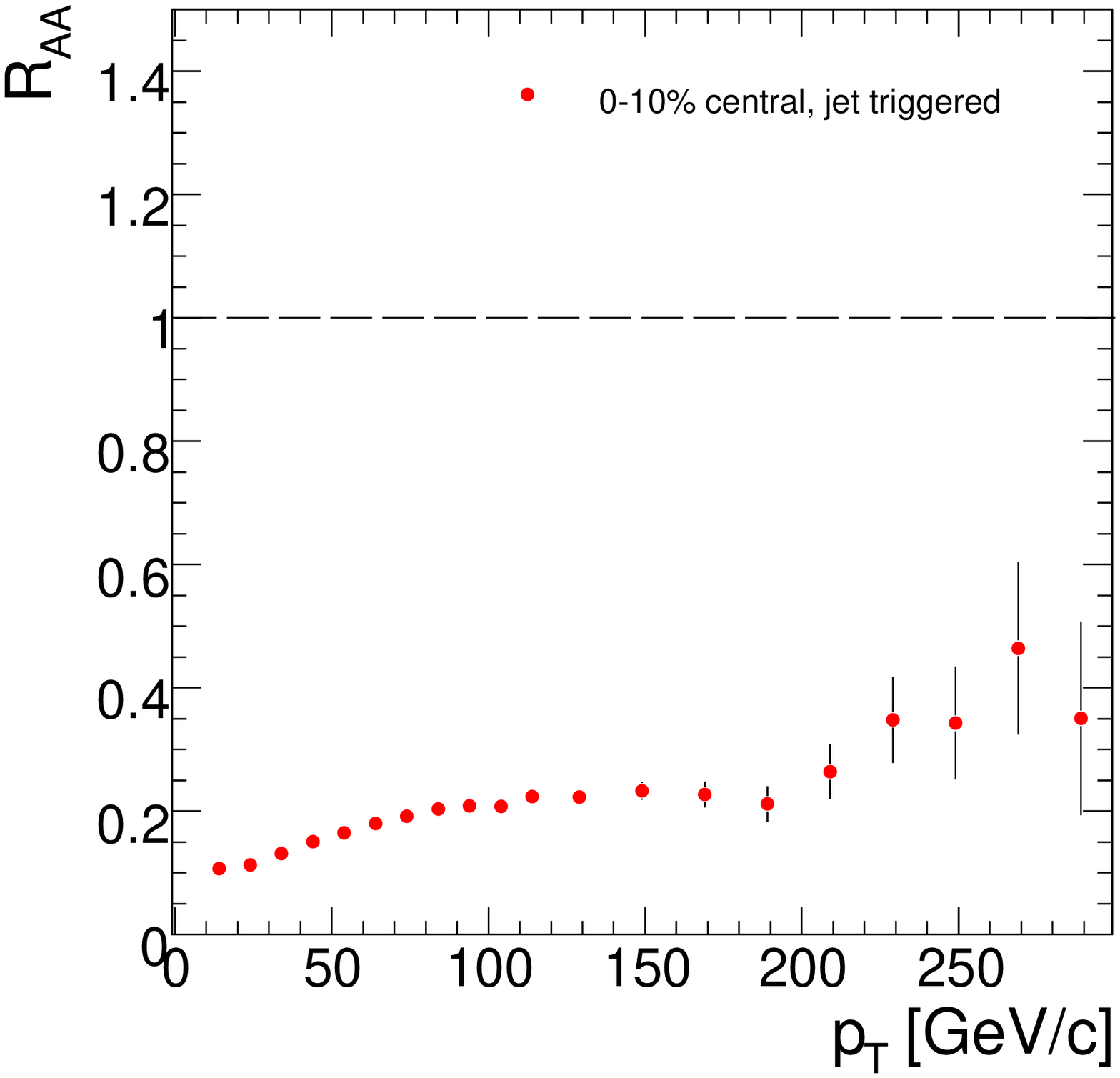}}
    \parbox[t]{0.47\textwidth}{\caption{Charged particle $R_{AA}(p_T)$ 
    for an integrated luminosity of 0.5 nb$^{-1}$ using the minbias sample.}
    \label{cms_minbias}}
    \hfill
    \parbox[t]{0.47\textwidth}{\caption{Same as Fig. 5  but for data triggered on high-$p_T$ jets.}
    \label{raa}}
\end{figure}

\begin{figure}[ht]
\centering
    \includegraphics[scale=0.47]{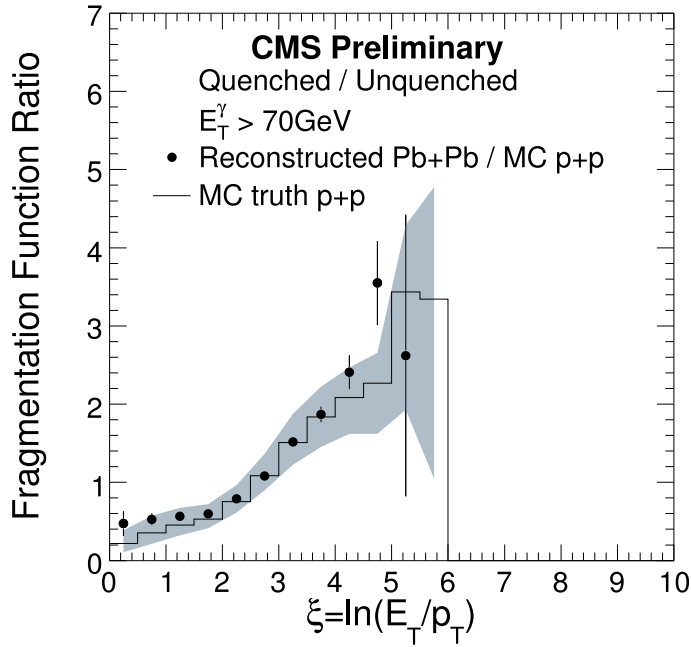}
    \caption{The ratio of the reconstructed quenched fragmentation function to the 
       unquenched one (filled circles) is compared with the Monte-Carlo truth (solid histograms) for the
       integrated luminosity of 0.5~nb$^{-1}$.
    \label{ff_cms}}
\end{figure}

\begin{figure}[ht]
\centering
    \includegraphics[scale=0.48]{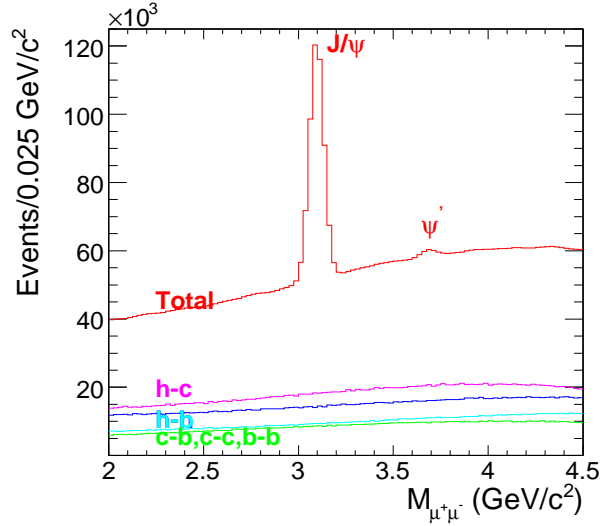}
    \caption{(color online) Invariant mass spectra of opposite--sign muon pairs in $J/\psi$ mass
    range with $dN_{ch}/d\eta|_{\eta=0} =$~2500 with both muons in $|\eta|<$~2.5. 
    \label{cms_jpsi}}
\end{figure}
\begin{figure}[h!t]
\centering
    \includegraphics[scale=0.48]{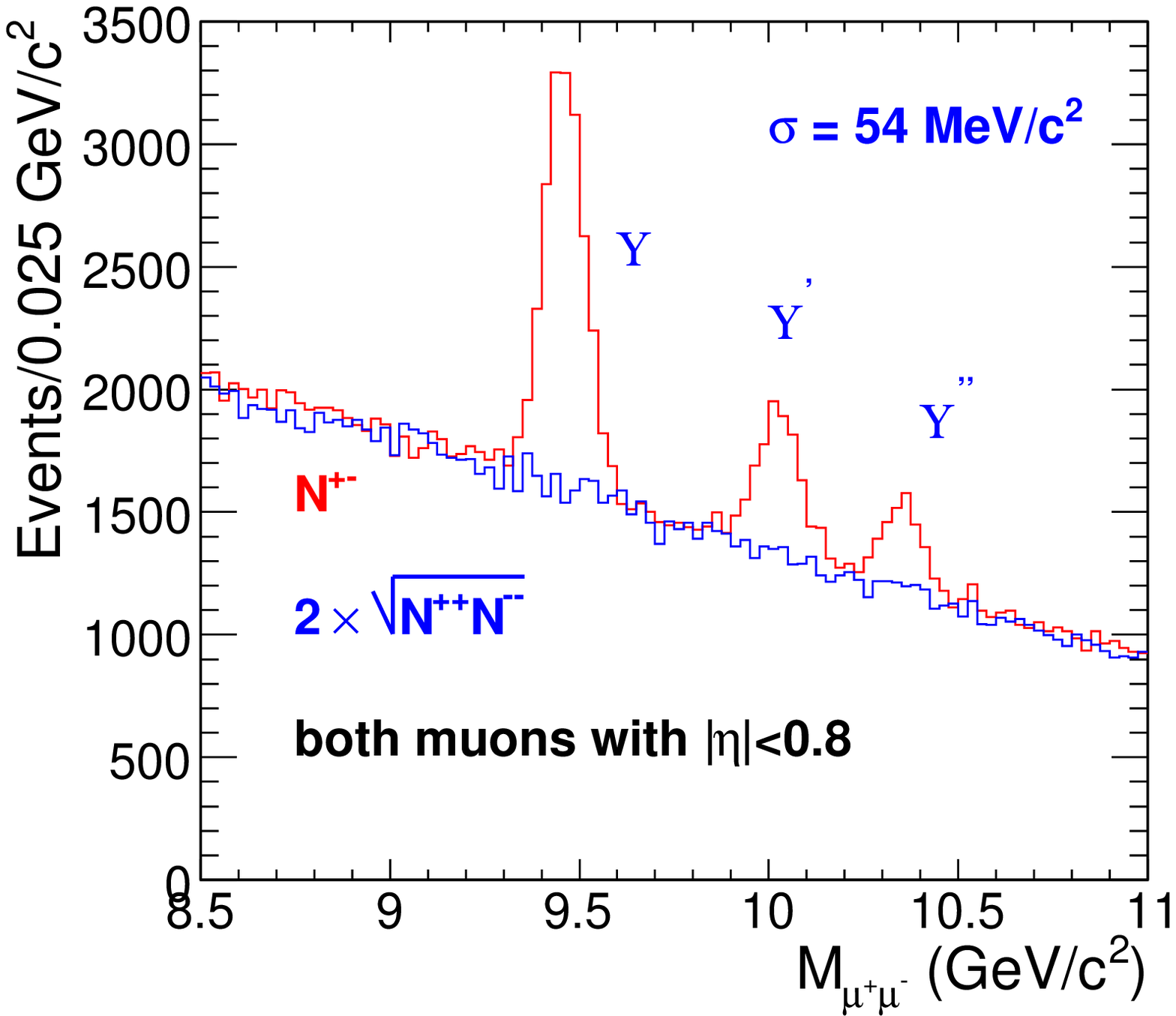} 
    \caption{(color online) Invariant mass spectra of $\mu^+\mu^-$ pairs in $\Upsilon$ mass
    range, $dN_{ch}/d\eta|_{\eta=0} =$~2500 and both muons have $|\eta|<$~0.8. 
    \label{cms_ups}}
\end{figure}

%    \resizebox{0.4\textwidth}{!}{\includegraphics{EPS/cms_event_plane.eps}}
%    \parbox[t]{0.47\textwidth}{\caption{The reaction plane resolution estimated with the 
%                                        electromagnetic calorimeter.}\label{evplane}}

\section{Hard ("tomographic") probes of dense QCD matter}
  Hard probes, i.e. particles with large transverse momentum and/or high mass are of crucial importance for 
several reasons: (i) they originate from parton scattering with large momentum transfer Q$^2$ and are directly
coupled to the fundamental QCD degrees of freedom; (ii) their production timescale is short, allowing them to
propagate and potentially be affected by the medium; (iii) their cross-sections can be theoretically predicted
with pQCD.
%\subsection{Jets and high-$p_T$ hadrons production}

One of the major discoveries at the RHIC is the  suppression of  high $p_T$ hadrons compared to what
would be expected based on the corresponding number of binary pp collisions. This effect is known as  jet quenching.  The nuclear modification factor, $R_{AA}(p_T)$  is defined by the ratio of particle yield in heavy-ion collisions to the binary collisions scaled yield in p+p collisions. It provides a convenient first measure of the strength of jet quenching. In a typical run without the high level trigger we expect to be able to measure $R_{AA}$ out to 150 GeV see  Fig.~\ref{cms_minbias}. Using the HLT we can double our $p_T$ reach, see  Fig.~\ref{raa} ~\cite{hiaddendum}. 

%~\ref{cms_minbias} but 
%    Expected statistical reach for the nuclear modification function, $R_{AA}(p_T)$ for inclusive charged hadrons
 %   in central Pb-Pb collisions generated with HYDJET~\cite{hydjet} for a nominal integrated luminosity of 0.5 nb$^{-1}$ 

   New hard probes are available at the LHC, such as boson-tagged ($\gamma$,$Z^0$) jet production. We have developed algorithms to reconstruct jets, high-$p_T$ tracks and photons  in the heavy ion environment ~\cite{hiaddendum}. Photon-jet events are a convenient way to study jet fragmentation since the photon does not lose energy as it propagates through the partonic medium. Figure 
    ~\ref{ff_cms} shows that the ratio of the reconstructed quenched fragmentation function to the 
unquenched one can be measured accurately for tracks with $p_T$ values between 1 and  7 * $10^{-3}$ times the momentum of the photon  ~\cite{cms_gammajet}.

%\subsection{Quarkonium production}

  The suppression of heavy-quark bound states in high energy A-A collisions was one of the first proposed 
  signatures for a deconfined medium of quarks and gluons to be actually observed in experiment 
 ~\cite{NA38,satz}. 
 At RHIC PHENIX has found that the suppression is stronger at forward rapidity, than at mid-rapidity~\cite{jpsi_rhic_1}. 
 Recent lattice calculations predict a 
step-wise suppression of the J/$\psi$ and $\Upsilon$ families because of the different melting temperatures for
each $Q\bar{Q}$ state~\cite{satz2}. At the LHC the $\Upsilon$ family will be available with large statistics for the
first time. Unlike the J/$\psi$ family the bottomonium family will be less affected by the recombination process since there are only a few b and bbar quarks per events. Our simulated  dimuon spectra for the J/$\psi$ and $\Upsilon$ families  ~\cite{hiaddendum} are shown in 
Figs.~\ref{cms_jpsi} and \ref{cms_ups}. The mass resolution for 
$\Upsilon$ is about 54 MeV/c$^2$ in the barrel and it worsens to 90 MeV/c$^2$ if the endcap detectors are included.
For the  J/$\psi$, our mass resolution is  35 MeV/c$^2$ for CMS in full $\eta$ range. Around 20 000 $\Upsilon$s and $\simeq$200, 000  J/$\psi$s are expected for an integrated luminosity of  0.5 nb$^{-1}$.

\section{Summary}
CMS is a superb detector for measuring muons, photons, jets and charged tracks from lead-lead collisions at high rate over a very large rapidity range.  In this paper we have been able to highlight some of our capabilities to use both  soft and hard probes such as 
multiplicity, low  and high $p_T$ spectra of charged particles, photons, jets and quarkonia to study QCD at very high temperatures, high energy densities and also very low x. Lack of space prevents us from discussing other probes such as dijet correlations and ultra-peripheral collisions.  One can think of CMS as a generic detector for heavy ions that is able to measure almost the full phase space of the collisions. The collaboration is ready and eager to make the measurements described above but also is on the lookout  for the completely unexpected.

%% end of main text

\section*{Acknowledgments} % please check/modify, comment out or delete if not needed
We wish to thank our CMS colleagues for help in presenting the CMS Heavy Ion program. We also gratefully acknowledges support from Russian Foundation for Basic Research, grant No 08-02-91001; the US National Science Foundation grant PHY-0449913;  and the US Department of Energy grant DE-FG03-96ER40981. 

 % do not change 
\end{document}